\documentclass[preprint,proceedings]{rmaa}

\usepackage{paralist}

\usepackage{psfrag,color}

\SetYear{2003}
\SetConfTitle{Compact Binaries in the Galaxy and Beyond}

\title{The mysterious Nature of HS\,2331+3905} 

\author{
  S. Araujo-Betancor\altaffilmark{1,2} 
  B.T. G\"ansicke\altaffilmark{3,2}
  H.-J. Hagen\altaffilmark{4}  
  T. Marsh\altaffilmark{3,2}
  J. Thorstensen\altaffilmark{5}
  E. Harlaftis\altaffilmark{6} 
  R.E. Fried\altaffilmark{7} 
  D. Engels\altaffilmark{4}  
}
\altaffiltext{1}{Space Telescope Science Institute, USA.}
\altaffiltext{2}{University of Southampton, UK.}
\altaffiltext{3}{University of Warwick, UK.}
\altaffiltext{4}{Hamburger Sternwarte, Germany.}
\altaffiltext{5}{Dartmouth College, USA.}
\altaffiltext{6}{Athens National Observatory, Greece.}
\altaffiltext{7}{Braeside Obsevatory, USA.}

\shortauthor{Araujo-Betancor et al.}
\shorttitle{The mysterious nature of HS\,2331}

\suppressfulladdresses
\listofauthors{S. Araujo-Betancor, B.T. G\"ansicke, H.-J. Hagen,
  T. Marsh, J. Thorstensen, E. Harlaftis, R.E. Fried \& D. Engels}
\indexauthor{Araujo-Betancor, S.}
\indexauthor{G\"ansicke, B. T.}
\indexauthor{Hagen, H.-J}
\indexauthor{Marsh, T.}
\indexauthor{Thorstensen, J.}
\indexauthor{Harlaftis, E.}
\indexauthor{Fried, R. E.}
\indexauthor{Engels, D.}

\abstract{We report the discovery of one unique cataclysmic variable
drawn from the Hamburg Quasar Survey, HS\,2331+3905. Follow-up
observations obtained over three years unveiled a very unusual
picture. The large amplitude 3.5\,h radial velocity variations
obtained from our optical spectroscopy is not the orbital period of
the system, as one would normally expect. Instead, extensive CCD
photometry strongly suggests that HS\,2331+3905 is a short orbital
period cataclysmic variable with $P_{\rm orb}= 81.09$\,min, containing
a cold white dwarf which appears to exhibit ZZ\,Ceti pulsations.}

\addkeyword{Stars: binaries:close}
\addkeyword{Stars: Cataclysmic Variables}
\addkeyword{Stars: individual: HS\,2331+3905}

\begin{document}
\maketitle
\section{Introduction}
HS\,2331+3905 (HS\,2331 thereafter) was selected as a cataclysmic
variable (CV) candidate on the basis of its spectral characteristics
in the Hamburg Quasar Survey (HQS; Hagen et al.~1995; G\"ansicke et
al. 2002). The identification spectrum of HS\,2331 contains broad
double-peaked Balmer emission lines, clear signs of the presence of an
accretion disc, flanked by extremely broad absorptions throughs,
indicating that this CV contains a relatively cold white dwarf. The
red part of the spectrum does not contain any spectral features that
could be ascribed to the emission of the secondary. Here we report
follow-up (ground and space) photometry and spectroscopy of
HS\,2331, obtained over a three year period after its identification.

\begin{figure*}[!t]
\hfill
\includegraphics[width=5.6cm, angle=270]{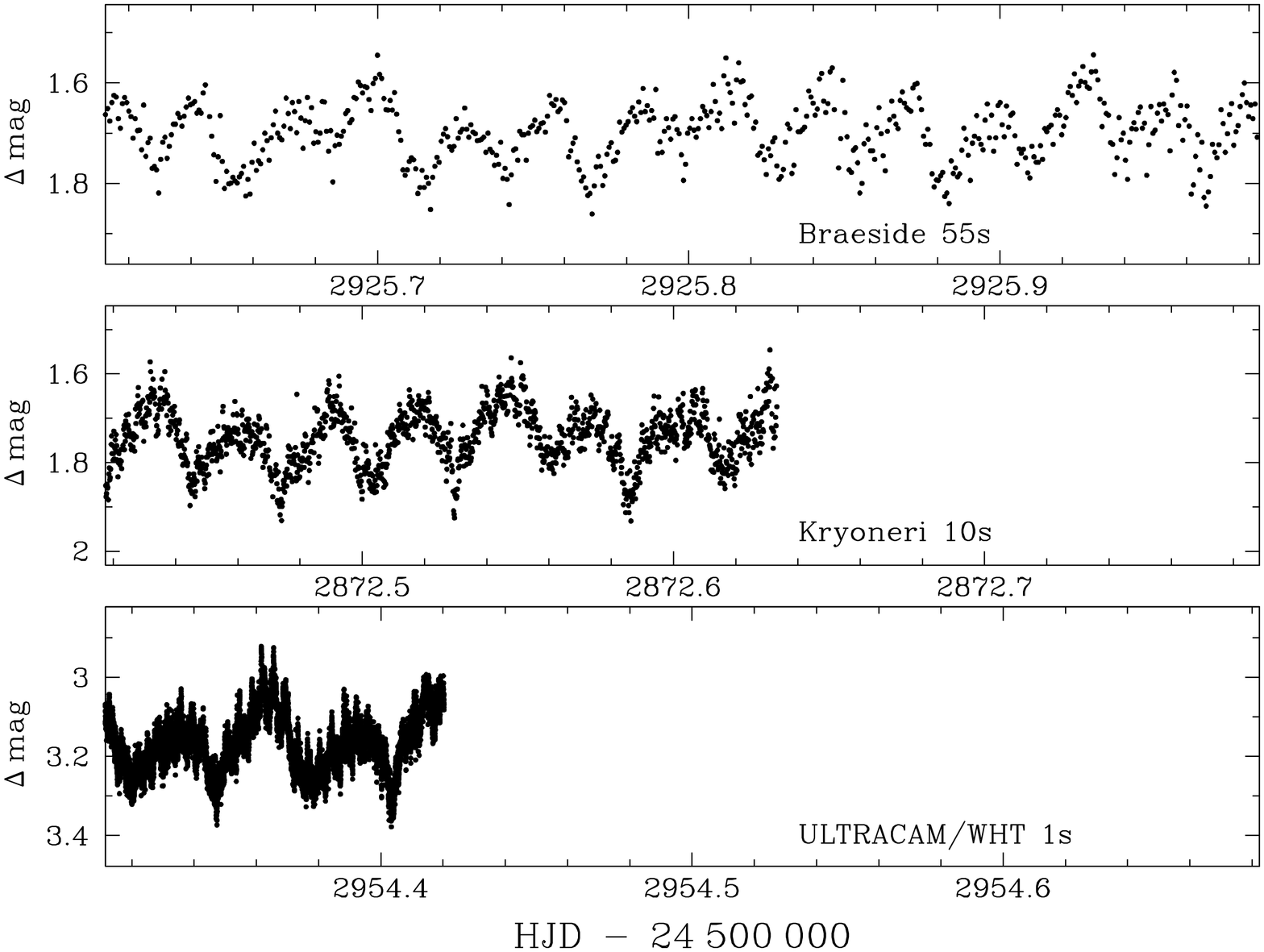}%
\hfill
\includegraphics[width=5.6cm, angle = 270]{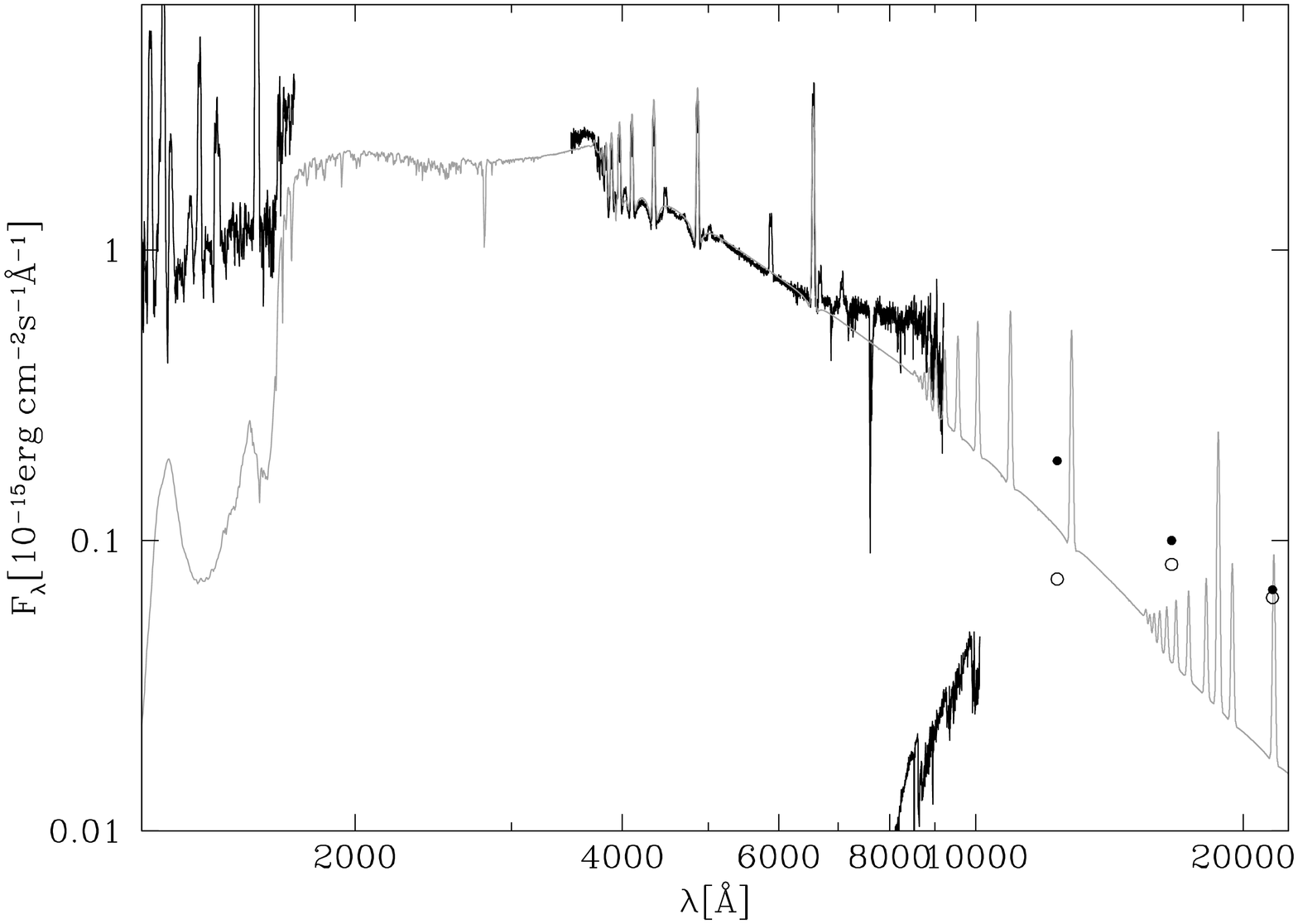}
\hfill
\caption{\textit{Left panel:} Samples of the light curves of
HS\,2331 obtained from differential CCD photometry. The names and
numbers indicate the observatory and time resolution used in each of
the observations respectively. \textit{Right panel:} Combination of
$FUV$, optical spectra and 2MASS colours of HS2331 (dark line and
filled circles) plotted with the best three-component model fit
(grey line and open circles). See text for details.}
\label{f-fig1}
\end{figure*}

\section{Observations}
The left panel of Fig.\,\ref{f-fig1} shows sample light curves of
HS\,2331 obtained from differential CCD photometry. The morphology of
the photometric modulation is best described by a double-humped
pattern with a period of $\sim 80$\,min, with narrow dips centred on
the observed minima between humps, which we identify as grazing
eclipses.  The higher time resolution data, in the middle and bottom
panel of the same figure, reveal additional variability on time scales
of 5\,min and 1\,min.  A period analysis of the entire CCD photometry
of HS\,2331 - more than 20\,000 data points - confirms the
multiperiodic variability directly seen in the light curves. The
likely orbital period derived from the period analysis, and ratified
by folding all the data over it, is $P_{\rm
orb}=81.0852\pm0.0002$\,min.  In addition to the orbital period, the
period analysis also reveals strong peaks at 83.38\,min, 5.61\,min and
1.12\,min.  The 83.38\,min is in the right frequency range to
represent a superhump period (i.e. 1--2$\%$ longer than $P_{\rm
orb}$). The power spectra around the 5.61\,min and 1.12\,min signals
show an extremely complex structure indicative of a superposition of
many frequencies. This type of structures are found in ZZ\,Ceti
pulsators which can be explained by a number of non-radial pulsations
modes, their harmonics, and various linear combinations of modes
(e.g. Kotak et al. 2002).

Radial velocity variations of the Balmer and He\,I emission lines
revealed yet another period at $\sim3.5$\,h. This period seems to
drift throughout the three years of observations, and we could not
identify a single period that will satisfactorily fold all the
available radial velocity data. We conclude that a persistent
large-amplitude radial velocity variation with a period $\sim 3.5$\,h
is present in HS\,2331, however, this variation is not coherent but
its period drifts on time scales of days.

In addition to optical spectroscopy we have obtained a
\textit{HST}STIS far-ultraviolet ($FUV$) spectrum of HS\,2331. The
$FUV$-optical spectra combined with the 2-MASS $JHK$ colours of
HS\,2331 allow us to confirm the presence of a low-temperature white
dwarf of $T_{\rm wd}\sim$ 11\,000\,K, and to constrain the spectral
type of the secondary to be later than M9, consistent with the short
orbital period of the system (see right panel of
Fig.\,\ref{f-fig1}). The accretion disc contribution to the spectral
energy distribution of HS\,2331 was matched with a 6500\,K isothermal
and isobaric slab and a surface density $1.81\times10^{-2}$\,g\,$\rm
cm^{-2}$. The distance of the system estimated from the white dwarf
model fit is $\sim$100\,pc, consistent with the large proper motion of
the star, $\mu = 0.14''$.

\section{Discussion}

We have discovered a short orbital period system, HS\,2331, as
part of the HQS quest for new CVs. The orbital period of HS\,2331,
$P_{\rm orb}=81.09$\,min, was primarily defined by the detection of
coherent eclipses. From our three years of photometric data, HS\,2331
appears to be a permanent superhumper with $P_{\rm
SH}=83.38$\,min. The light curves of HS\,2331 display double-humps
with a period that is exactly half the orbital period (evident from a
direct inspection of the light curves in the left panel of
Fig.\,\ref{f-fig1}), suggesting that we are seeing some sort of
symmetric structure, such as e.g. two bright spots.  In addition,
HS\,2331, exhibits the photometric behaviour typical of ZZ\,Ceti
pulsators, showing multifrequency variability in the range
$\sim 60$\,s to $\sim300$\,s. The white dwarf temperature derived from
our fit, 11\,000\,K, is well within the instability strip for
ZZ\,Ceti pulsators.  In order to disentangle the multiperiodic
signature of the likely white dwarf pulsator in HS\,2331, we need to
organize a multi-site observing campaign to obtain long, continuous
stretches of high time resolution photometry.

All in all, the pieces of the jigsaw seems to come together, and we
are beginning to understand the nature of HS\,2331. There are
nevertheless, several points that we still need to address. The fact
that the dominant radial velocity variability does not correspond to
the orbital period of the system is particularly disconcerting. At
present, we have no explanation for this phenomenon, nor for the
physical significance of the 3.5\,hr radial velocity period. We are
not aware of any other system suffering from this problem, but the
reason for this may be that periods determined from radial velocity
studies are usually adopted unquestioned as reflecting the
corresponding orbital periods.

\end{document}